\documentclass[fleqn,12pt]{article}
\usepackage{epsfig}
\begin{document}
\textheight 22cm
\textwidth 15cm
\noindent
{\Large \bf Effects of parallel ion motion on zonal flow generation in ion-temperature-gradient mode turbulence}
\newline
\newline
J. Anderson\footnote{anderson.johan@gmail.com}, J. Li, Y. Kishimoto
\newline
Department of Fundamental Energy Science
\newline
Graduate School of Energy Science, Kyoto University, Gokasho, Uji, Kyoto 611-0011
\newline
\newline
\begin{abstract}
\noindent
The role of parallel ion motion for zonal flow generation in ion-temperature-gradient (ITG) mode turbulence is investigated with focus on the effects of acoustic modes and toroidicity on the zonal flow. One possible reason for the weak suppression of ITG turbulence by zonal flows found in experiments in the Columbia Linear Machine (CLM) [Phys. Plasmas {\bf 13} 055905 (2006)] might be due to the small toroidicity ($\epsilon_n = 2L_n/R$) in the experiment. The zonal flow is often directly dependent on the ITG mode and the coupling of zonal flow to acoustic modes and hence is directly affected by any change of the relevant parameters. The model consists of the continuity, temperature and parallel ion momentum equations for the ITG turbulence. The zonal flow time evolution is described by a Hasegawa-Mima like equation and a fifth order zonal flow dispersion relation is derived. The results are interpreted in terms of quality of zonal flows, i.e., the ratio of growth rate and real frequency ($Q = \Omega^{IM}/\Omega^{RE}$). It is found that the quality of the zonal flow rapidly decreases with decreasing toroidicity.
\end{abstract}
\newpage
\renewcommand{\thesection}{\Roman{section}}
\section{Introduction}
\indent
In recent experiments in the Columbia Linear Machine (CLM)~\cite{a10} it has been shown that the turbulence suppression of zonal flows is rather weak~\cite{a19}. The suppression of drift wave turbulence by quasi stationary zonal flows has been a major topic during the past years in magnetic fusion research. The zonal flows are poloidally and toroidally symmetric ($k_{\theta} = 0$, $k_{\parallel} = 0$) and radially inhomogeneous ($k_r \neq 0$) flow structures in toroidal plasmas~\cite{a11}-~\cite{a13}. Zonal flows are widely believed to be the main stabilizing mechanism for turbulence and one of the main keys to gain access to high confinement regimes in present and future magnetic confinement fusion devices such as ITER~\cite{a131}. The generation of zonal flows from drift wave turbulence has been extensively studied analytically~\cite{a11},~\cite{a14}-~\cite{a26} and in computer simulations using gyrokinetic~\cite{a35}-~\cite{a38} and advanced fluid models~\cite{a39}-~\cite{a43}. 

The Ion-Temperature-Gradient (ITG) driven mode is believed to be responsible for the anomalously high heat transport in fusion devices and it has been shown that it is one of the main ingredients in generation of zonal flows~\cite{a23}-~\cite{a26}. The ITG mode has by now been investigated in many different contexts and situations~\cite{a28}-~\cite{a442}. Anisotropic effects in perpendicular and parallel ion temperature gradient are out of the scope of the present analysis for the zonal flow generation as well as for the linear state~\cite{a441}. The main features of the ITG mode is captured by the more simple model with isotropic ion temperature and temperature gradient~\cite{a443}. Other features often found in the CLM experiment, e.g impurities, do not change the ITG mode qualitatively, however they may contribute though an impurity driven mode~\cite{a444} that in turn may drive a zonal flow. It has also been show that the direct interaction of $\vec{E} \times \vec{B}$ flows with zonal flows are rather weak~\cite{a21},~\cite{a26}.

In this particular paper both the slab like and toroidal branch of the ITG mode will be considered as the generating mechanism for the zonal flow. Here, the ITG mode physics is based on the equations for ion continuity, ion temperature and parallel ion motion~\cite{a29}-~\cite{a44}. The system is closed by considering Boltzmann distributed electrons and using the quasi neutrality condition.

The secondary generated zonal flow time evolution is governed by a Hasegawa-Mima like equation. The zonal flow excitation is studied through coupling of coherent modes. The drift wave is considered to be a monochromatic pump wave which may couple to the zonal flow through symmetric sidebands and this gives a system of four coupled waves~\cite{a15}-~\cite{a21}. The corresponding system may be reduced to a fifth order dispersion relation for the zonal flow. It was shown in Ref.~\cite{a20} that the coherent mode coupling models agree well with the corresponding model derived using the wave kinetic equation approach.

To estimate the zonal flow growth rate and real frequency the mode coupling saturation~\cite{a45} level is employed. In Ref.~\cite{a372} simulations of ITG modes in CLM were performed and a good agreement between the simulations and the experiments were found on the saturation level. This particular finding validates the use of the mode coupling saturation level used in this study.

However, the efforts of comparing analytical estimations with measurements are rather limited. The effects of coupling between the acoustic modes and the zonal flows are mostly unexplored as well. It has been indicated in gyrokinetic turbulence simulations that small density gradient scale lengths ($L_n$) increases the transport~\cite{a371} which encourages the present theoretical study. 
These issues will be discussed in the present study with an emphasis on the differences in the generation from the slab like and toroidal ITG mode branch. 

The results are also interpreted in terms of quality of zonal flows i.e. the ratio of zonal flow growth rate and real frequency ($\Omega^{IM}/\Omega^{RE}$)~\cite{a26}. The toroidally driven zonal flows exhibit a much larger ratio of growth rate and frequency compared to slab like driven flows (factor of three difference). The slab ITG generated zonal flow is only weakly dependent on $\eta_i$. Moreover, the quality of zonal flows is rather insensitive to $\eta_i$, in contrast to the results reported using the Wave Kinetic Equation (WKE) modeling. In many aspects the results found comparing the WKE modeling and the coherent mode coupling are qualitatively similar~\cite{a20}. In the WKE modeling a strongly decreasing zonal flow quality with increasing $\eta_i$ was found~\cite{a26}, however in the present model the decrease of zonal flow quality is visible for larger $\eta_i$. It is also important to note that the difference in quality for slab like and toroidally generated zonal flows is approximately a factor of three. It is indicated that there is a transition from a state with stationary zonal flows to a state with oscillatory zonal flows. In all parameter scalings the coherent mode coupling model agrees well with a model derived using the WKE model for the same background physics. 
  
The paper is organized as follows. In section II the toroidal and slab ITG driven modes are presented and in section III the equations for the zonal flow are discussed. Section IV, contains results and discussion and finally there is a summary in section V.

\section{Toroidal and slab ion-temperature-gradient driven modes}
The model for the ITG driven modes consists of the ion continuity, ion temperature and the parallel ion momentum equations~\cite{a29}-~\cite{a44}. The system of equations is closed by using quasi neutrality with Boltzmann distributed electrons. The effects of magnetic shear and finite beta are neglected in the present study. It has been found earlier that the effect of parallel ion motion on the ITG mode growth rate is rather weak~\cite{a28}. It has recently been found that magnetic shear may modify the zonal flow generation significantly for the Trapped Electron Modes (TEM), however, the effects of magnetic shear on the ITG mode are modest and are also out of the scope of the present paper~\cite{a25}. In this section the linear regime of the ITG mode will be discussed and the basic equations are given by; 
 
\begin{eqnarray}
\frac{\partial \tilde{n}}{\partial t} - \left(\frac{\partial}{\partial t} - \alpha_i \frac{\partial}{\partial y}\right)\nabla^2_{\perp} \tilde{\phi} + \frac{\partial \tilde{\phi}}{\partial y} - \epsilon_n g \frac{\partial}{\partial y} \left(\tilde{\phi} + \tau \left(\tilde{n} + \tilde{T}_i \right) \right) + \frac{\partial \tilde{v}_{i ||}}{\partial z} = \nonumber \\
- \left[\phi,n \right] + \left[\phi, \nabla^2_{\perp} \phi \right] + \tau \left[\phi, \nabla^2_{\perp} \left( n + T_i\right) \right] \\
\frac{\partial \tilde{T}_i}{\partial t} - \frac{5}{3} \tau \epsilon_n g \frac{\partial \tilde{T}_i}{\partial y} + \left( \eta_i - \frac{2}{3}\right)\frac{\partial \tilde{\phi}}{\partial y} - \frac{2}{3} \frac{\partial \tilde{n}}{\partial t} = \nonumber \\
- \left[\phi,T_i \right] + \frac{2}{3} \left[\phi,n \right] \\
\frac{\partial \tilde{v}_{i ||}}{\partial t} = - (\frac{\partial \tilde{\phi}}{\partial z} + \tau \frac{\partial (\tilde{n} + \tilde{T_i})}{\partial z}) - \left[ \tilde{\phi}, \tilde{v}_{i ||} \right].
\end{eqnarray}

Here $\left[ A ,B \right] = (\partial A/\partial x) (\partial B/\partial y) - (\partial A/\partial y) (\partial B/\partial x)$ is the Poisson bracket with the additional definitions $\tilde{n} = \delta n / n_0$, $\tilde{\phi} = e \delta \phi /T_e$, $\tilde{T}_i = \delta T_i / T_{i0}$ as the normalized ion particle density, the electrostatic potential and the ion temperature respectively. In the forthcoming equations $\tau = T_i/T_e$, $\vec{v}_{\star} = \rho_s c_s \vec{y}/L_n $, $\rho_s = c_s/\Omega_{ci}$ where $c_s=\sqrt{T_e/m_i}$, $\Omega_{ci} = eB/m_i c$. We also define $L_f = - \left( d ln f / dr\right)^{-1}$, $\eta_i = L_n / L_{T_i}$, $\epsilon_n = 2 L_n / R$ where $R$ is the major radius and $\alpha_i = \tau \left( 1 + \eta_i\right)$. The perturbed variables are normalized with the additional definitions $\tilde{n} = (L_n/\rho_s) \delta n / n_0$, $\tilde{\phi} = (L_n/\rho_s e) \delta \phi /T_e$, $\tilde{T}_i = (L_n/\rho_s) \delta T_i / T_{i0}$, $\tilde{v}_{i ||} = (L_n/\rho_s) \delta v_{i ||}/c_s$ as the normalized ion particle density, the electrostatic potential, the ion temperature and the parallel ion velocity respectively. The perpendicular length scale and time are normalized to $\rho_s$ and $L_n/c_s$, respectively. The geometrical quantities are calculated in the strong ballooning limit ($\theta = 0 $, $g_i\left(\theta = 0, \kappa \right) = 1/\kappa$ where $g_i\left( \theta \right)$ is defined by $\omega_D \left( \theta \right) = \omega_{\star} \epsilon_n g_i\left(\theta \right)$)~\cite{a31},~\cite{a44}. The effects of safety factor and magnetic shear can also be accommodated by the factor $g_i$~\cite{a23}.
The linear dispersion relation coming from Eqs 1, 2 and 3 is,
\begin{eqnarray}
\omega^3 \left(1+k_{\perp}^2\right) + \omega^2 k_y \left(-(1-(1+\frac{10}{3}\tau)\epsilon_n g_i) + (\alpha_i  + \frac{5}{3}\tau \epsilon_n g_i)k_{\perp}^2 \right) & + & \nonumber \\
\omega\left( \tau \epsilon_n g_i k_y^2(\eta_i - \frac{7}{3} + \frac{5}{3}(1+\tau)\epsilon_n g_i + \frac{5}{3}k_{\perp}^2\alpha_i )- k_{\parallel}^2 (1 +  \frac{5}{3} \tau) \right) & - & \nonumber \\
k_{\parallel}^2  k_y \tau \left( (\eta_i -  \frac{2}{3}) +  \frac{5}{3} (1+\tau)\epsilon_n g_i \right) & = & 0. \nonumber \\
\end{eqnarray}
Here the fields for the linear ITG mode are expanded according to $\tilde{\xi}_{DW}(x,y,t)  =  \xi_{DW} e^{i(k_x x + k_y y + k_{\parallel} z - \omega t)} + c.c.$ for ($\tilde{n}_i,\tilde{\phi},\tilde{T}_i, \tilde{v}_{i ||}$). The dispersion relation for the linear ITG mode including parallel ion motions effects is of 3rd order, however if $k_{\parallel}^2$ effects are neglected the dispersion relation is transformed into the previous result~\cite{a23}. The effects of parallel ion motion on the real frequency and the growth rates are small. The important differences are that the real frequencies and growth rates are significantly changed for small toroidicity and in addition the $\eta_i$-threshold is decreased. In this particular case it is of interest to monitor the zonal flow generation from toroidal and slab like ITG mode turbulence. In the limit of small $\epsilon_n=2 L_n/R$ and small Finite Larmor Radius (FLR) effects an approximate solution may be found, $\omega_{Slab}^2 \approx - k_{\parallel}^2 \tau (\eta_i - 2/3)$, in analogy with Ref.~\cite{a27}. In comparison with the model in Ref.~\cite{a27}, the present model yields a positive threshold value $\eta_{i th} = 2/3$ for the slab like ITG mode. In the toroidal ITG mode branch approximate solutions are close to those found without parallel ion motion e.g. Eqs 8 - 10 in Ref.~\cite{a23} (given here for convenience),
\begin{eqnarray}
\omega_r & \approx & \frac{k_y}{2\left( 1 + k_{\perp}^2\right)} \left( 1 - \left(1 + \frac{10\tau}{3} \right) \epsilon_n g_i - k_{\perp}^2 \left( \alpha_i + \frac{5}{3} \tau \epsilon_n g_i \right)\right)  \\
\gamma & \approx & \frac{k_y}{1 + k_{\perp}^2} \sqrt{\tau \epsilon_n g_i\left( \eta_i - \eta_{i th}\right)} \\
\eta_{i th} & \approx & \frac{2}{3} - \frac{1}{2 \tau} + \frac{1}{4 \tau \epsilon_n g_i} + \epsilon_n g_i\left( \frac{1}{4 \tau} + \frac{10}{9 \tau}\right)
\end{eqnarray}
where $\omega = \omega_r + i \gamma$.
The origins of the slab and toroidal ITG mode instabilities are inherently different: the slab like branch is driven unstable by parallel acoustic waves; while the other branch is driven by toroidicity~\cite{a27}.
   
\section{The model for zonal flow generation}
 In this section the derivation of the zonal flow dispersion relation will be explained briefly. The method used has been covered extensively in previous papers~\cite{a11},~\cite{a15}-~\cite{a21}. In this paper the effects on the zonal flow growth rate and real frequency of toroidicity and the coupling to the acoustic modes will be investigated. There are some limitations to this derivation of the zonal flow generation namely the non-linear terms in the equation for temperature ($[\phi, T_i] = 0$) and in the parallel ion momentum equation are neglected ($[\phi, v_{i ||}] = 0$). 
The procedure is to start with a monochromatic pump wave (defined above) that drives the zonal flow through a modulational coupling via the sidebands. The fields for the zonal flow and the sidebands are expanded as follows,
\begin{eqnarray}
\tilde{\xi}_{ZF}(x,y,t) & = & \xi_{ZF} e^{i(q_x x - \Omega t)} + c.c. \\
\tilde{\xi}_\pm(x,y,t) & = & \xi_{\pm}e^{i((q_x \pm k_x)x \pm k_y y \pm k_{\parallel} z - \omega_{\pm }t)} + c.c.
\end{eqnarray}
The secondary generated zonal flow is defined as $\tilde{\xi}_{ZF}$ ($\tilde{\phi}, \tilde{T}_i$ remembering that the zonal flow does not have a density component), $\tilde{\xi}_\pm$ are the possible sidebands ($\tilde{n}_i, \tilde{\phi}, \tilde{T}_i$). The radial, poloidal and parallel wave numbers are $(k_x,k_y,k_{\parallel})$, $\omega$ is the complex frequency of the drift wave, $q_x$ is the radial wave number, $\Omega$ is the complex frequency of the zonal flow. The frequencies couple as $\omega_{\pm} = \Omega \pm \omega$. 

The equation for the time evolution for the zonal flow, that is assumed to be Hasegawa-Mima type, is expanded using the above fields,
\begin{eqnarray}
-\frac{\partial}{\partial t} \nabla_x^2 \phi_{ZF} -\mu \nabla_x^4 \phi_{ZF} = \left[\phi, \nabla_{\perp}^{2} \phi\right].
\end{eqnarray}

In addition the sidebands are to be determined. The sidebands evolve according to Eq. 1 and also identify the coupling between the drift wave and the zonal flow as the sideband component. This gives a problem with four coupled waves where the zonal flow seed may be unstable and grow exponentially. This gives the equation system,
\begin{eqnarray} 
(i\Omega q_x^2 - \mu q_x^4) \phi_{ZF} & = & k_y q_x^2 [a_{+} \phi_{DW}^{*} \phi_{+} + a_{-}\phi_{DW} \phi_{-}] \\
\phi_{+} & = &  i\beta q_x k_y k_{\perp -}\omega_{+}\frac{\phi_{DW} \phi_{ZF}}{\omega_{+}(\omega_{+}-k_y+\epsilon_n g_i \beta k_y) - k_{\parallel}^2 \beta} \\
\phi_{-} & = & - i\beta q_x k_y k_{\perp -}\omega_{-}\frac{\phi_{DW}^{*} \phi_{ZF}}{\omega_{-}(\omega_{-} + k_y - \epsilon_n g_i \beta k_y) - k_{\parallel}^2 \beta}.
\end{eqnarray}
In deriving these equations the FLR effects are considered to be small. Here 
\begin{eqnarray} 
a_{\pm} & = & 2 k_x \pm q_x \\
k_{\perp -} & = & k_x^2 + k_y^2 - q_x^2 \\
\beta & = & 1 + \tau + \tau \delta \\
\delta & = & \frac{(\eta_i - \frac{2}{3})k_y + \frac{2}{3}\omega}{\omega + \frac{5}{3}\tau \epsilon_n g k_y}.
\end{eqnarray}
It should be observed that if $k_{\parallel} \rightarrow 0$, Eqs. 11 - 13 are reduced to the corresponding equations derived without parallel ion motion~\cite{a20}. It is also important to observe the fact that at $q_x = 0$ the Reynolds stress term vanishes and the drive for the secondary generated zonal flow disappears. Using Eqs. 11 - 13 it is now possible to derive a fifth order dispersion relation for the zonal flow component. 
\begin{eqnarray}
(\Omega + i \mu q_x^2) \bar{\Omega}_{+} \bar{\Omega}_{-} &  = &  [ K_{+}(\Omega + \omega) \bar{\Omega}_{-} + K_{-}(\Omega - \omega)\bar{\Omega}_{+} ] |\phi_{DW}|^2 \\
\bar{\Omega}_{+} & = & (\Omega + \omega)(\Omega +\alpha) - k_{\parallel}^2 \beta \\
\bar{\Omega}_{-} & = & (\Omega - \omega)(\Omega - \alpha) - k_{\parallel}^2 \beta\\
K_{+} & = & k_y k_{\perp -} \beta a_{+}\\
K_{-} & = & k_y k_{\perp -} \beta a_{-}\\
\alpha & = & \omega - k_y + \epsilon_n g_i k_y \beta
\end{eqnarray}

A detailed study of the dispersion relation shows the explicit effect of parallel ion motion on the zonal flow. Expanding the dispersion relation around the solution $\Omega_0$ for $k_{\parallel} = 0$, where the dispersion relation is reduced to third order, with $\Omega = \Omega_0 + \delta \Omega$ and solving for $\delta \Omega$,
\begin{eqnarray}
\Im (\delta \Omega) \approx \frac{k_{\parallel}^2 \beta}{5 (\Omega_0^{IM})^4}(K_{-}-K_{+})\gamma |\phi_{DW}|^2 \\
\Re (\delta \Omega) \approx \frac{k_{\parallel}^2 \beta}{5 (\Omega_0^{IM})^4}(K_{-}-K_{+})\omega_{r} |\phi_{DW}|^2.
\end{eqnarray}
Here it is assumed that the zonal flow growth rate is much larger than the real frequency ($\Omega_0^{IM} >> \Omega_0^{RE}$). The important observation from Eqs. 24 - 25 is the rapid variation with the parallel wave number in the change in the zonal flow growth rate and the real frequency in the case of slab like ITG mode generated zonal flows. The slab like generated zonal flow growth rate and real frequency change in the same manner as the slab ITG mode. This also means that the zonal flow is driven by the acoustic waves through the slab like ITG mode instability. The Eqs. 24 - 25 also indicates a weak dependency of the parallel wave number on $\delta \Omega$ for the zonal flows generated from toroidal ITG mode turbulence.   

The fifth order dispersion relation (Eq. 18) is solved numerically and special attention is given to the behavior of zonal flow generation as a function of $\eta_i$, $\epsilon_n$ and $k_{\parallel}$. The results are also contrasted to previous results obtained in models excluding parallel ion motion.

\section{Results and discussion}
 In the present section the solutions to the zonal flow dispersion relation are discussed and compared with previous findings. The zonal flow dispersion relation Eq. 18 is solved numerically. In the numerical analysis it is assumed that the mode coupling saturation level is reached for the drift wave turbulence~\cite{a45}
\begin{eqnarray}
\tilde{\phi}_{DW} = \frac{\gamma}{\omega_{\star}}\frac{1}{k_y L_n}.
\end{eqnarray}
Of particular interest in the present work is the effect of toroidicity on the zonal flow generation. In this analysis this effect is included in the parameter $\epsilon_n$ where the slab like ITG mode is found for $\epsilon_n = 0$. The results are presented both in direct scalings and as the quality of zonal flows. The quality of zonal flows is defined as the ratio of the growth rate and the real frequency ($Q = \Omega^{IM}/\Omega^{RE}$).

First the effects of $\epsilon_n$ on zonal flow generation will be discussed. The quality of the zonal flow (Figure 1) with the parallel wave number ($k_{\parallel}^2$) as a parameter is investigated. The parameters are $\tau = 1$, $k_x = k_y = q_x = 0.3$, $\mu = 0$, $\eta_i = 4.0$, $k_{\parallel}^2 = 0.1$ (asterisk) and $k_{\parallel}^2 = 0.0$ (squares). Shown in Figure 1 is also the corresponding quality for the ITG mode $k_{\parallel}^2 = 0.1$ (plus). The parameters used in this paper are chosen as realistic values of a tokamak and for easy comparison of the results. The Figure shows a rapidly increasing quality of zonal flows with increasing $\epsilon_n = 2L_n/R$. A large negative value of the quality is more beneficial. The qualitative behavior of the model in a variation of the parallel wave number is similar except that for small or vanishing $\epsilon_n$ the slab-like mode generated zonal flows are captured. This is clearly indicating a stronger suppression effect of toroidally generated zonal flows. This result is to be compared with the absolute values of zonal flow growth rate and real frequency shown in Figure 2 in Ref.~\cite{a20}. The quality of the ITG mode shows a significantly different qualitative behavior (opposite trend) than the corresponding values for the zonal flow.

\begin{figure}
  \includegraphics[height=.3\textheight]{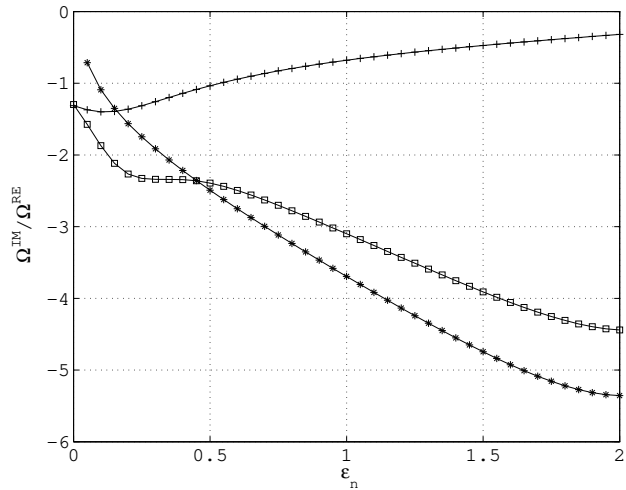}
  \caption{The zonal flow quality ($\Omega^{IM}/\Omega^{RE}$) as a function of $\epsilon_n$ with the parallel wave number ($k_{\parallel}^2$) as a parameter is displayed. The parameters are $\tau = 1$, $k_x = k_y = q_x = 0.3$, $\mu = 0$, $\eta_i = 4.0$, $k_{\parallel}^2 = 0.1$ (asterisk) and $k_{\parallel}^2 = 0.0$ (squares). The corresponding quality for the ITG mode is also shown $k_{\parallel}^2 = 0.1$ (plus).}
\end{figure}

Second, the slab and toroidally generated zonal flow qualities as a function of $\eta_i$ are compared with the parallel wave number ($k_{\parallel}^2$) as a parameter in Figure 2. The parameters are $\tau = 1$, $k_x = k_y = q_x = 0.3$, $\mu = 0$; $\epsilon_n = 1.0$ and $k_{\parallel}^2 = 0.0$ (asterisk), $k_{\parallel}^2 = 0.1$ (square); $\epsilon_n = 0.0$ and $k_{\parallel}^2 = 0.1$ (diamond). The figure shows that the quality of zonal flows is qualitatively rather insensitive to $\eta_i$, in contrast to the results reported in Ref.~\cite{a26}. The model in Ref.~\cite{a26} is derived using the Wave Kinetic Equation (WKE) and the results are approximately valid in the long wave length limit ($k_{\perp}^2 << 1$ and $k_{\parallel}^2 << 1$). In many aspects the results found comparing the WKE modeling and the coherent mode coupling are qualitatively similar~\cite{a20},~\cite{a23}-~\cite{a26}. In the WKE modeling a strongly decreasing zonal flow quality with increasing $\eta_i$ was found, however in the present model the decrease of zonal flow quality is visible for larger $\eta_i$ (not shown in the picture). It is also important to note that the difference in quality for slab like and toroidally generated zonal flows are approximately a factor of three. The results are shown for values well above the linear threshold ($\eta_{ith \: slab} = 2/3$ and $\eta_{ith \: tor} \approx 1.8$).

\begin{figure}
  \includegraphics[height=.3\textheight]{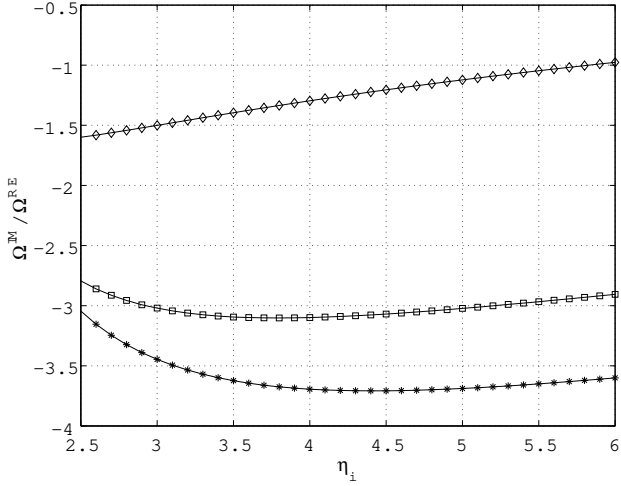}
  \caption{The zonal flow quality ($\Omega^{IM}/\Omega^{RE}$) as a function of $\eta_i$ for slab like and toroidally generated zonal flows is shown. The parameters are $\tau = 1$, $k_x = k_y = q_x = 0.3$, $\mu = 0$; $\epsilon_n = 1.0$ and $k_{\parallel}^2 = 0.0$ (asterisk), $k_{\parallel}^2 = 0.1$ (square); $\epsilon_n = 0.0$ and $k_{\parallel}^2 = 0.1$ (diamond).}
\end{figure}

In Figure 3 the zonal flow growth rate (positive) and real frequency (negative) (normalized to the ITG mode growth) as a function of $\eta_i$ with $\epsilon_n = 0.0$ (diamond) and $\epsilon_n = 1.0$ (square) are displayed. The other parameters are $\tau = 1$, $k_x = k_y = q_x = 0.3$, $\mu = 0$ and $k_{\parallel}^2 = 0.1$. The figure shows that the zonal flow is exhibiting the same behavior as the linear growth rate in scaling with $\eta_i$. For all values of $\eta_i$ the toroidally generated zonal flows have $\Omega^{IM}/\gamma_{ITG}>1$, whereas the slab like generated zonal flows have $\Omega^{IM}/\gamma_{ITG} \approx 1$. The modulus of real frequency ($|\Omega^{RE}|$) of the slab like generated zonal flows are, in comparison, much larger than the values from the corresponding toroidally generated zonal flow real frequency. The results support the findings that turbulence suppression is small in linear devices with small curvature (in the present paper represented with small $\epsilon_n$).

\begin{figure}
  \includegraphics[height=.3\textheight]{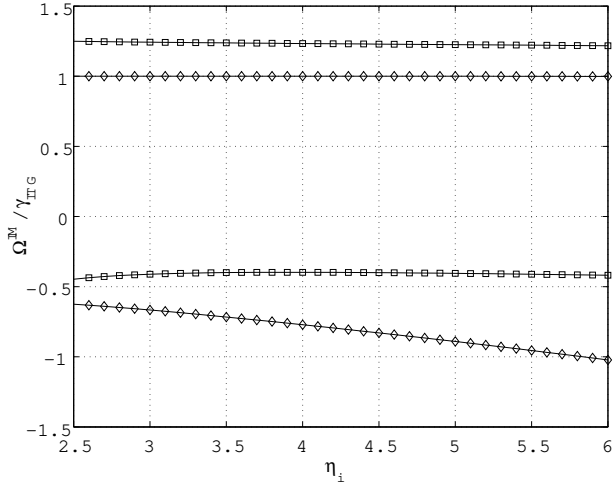}
  \caption{The zonal flow growth rate (positive) and real frequency (negative) (normalized to the ITG mode growth) as a function of $\eta_i$ with $\epsilon_n = 0.0$ (diamond) and $\epsilon_n = 1.0$ (square) are displayed. The other parameters are $\tau = 1$, $k_x = k_y = q_x = 0.3$, $\mu = 0$ and $k_{\parallel}^2 = 0.1$.}
\end{figure}

Third, the scaling of the zonal flow quality ($\Omega^{IM}/\Omega^{RE}$) as a function of parallel wave number ($k_{\parallel}^2$) is displayed in Figure 4. In the figure it is shown that the quality for the slab like generated zonal flows is quite insensitive to a change in the parallel wave number. This is due to a similar behavior of zonal flow growth rate and frequency with parallel wave number e.g. Figure 5. The quality of the toroidally generated zonal flows improves with increasing parallel wave number.

\begin{figure}
  \includegraphics[height=.3\textheight]{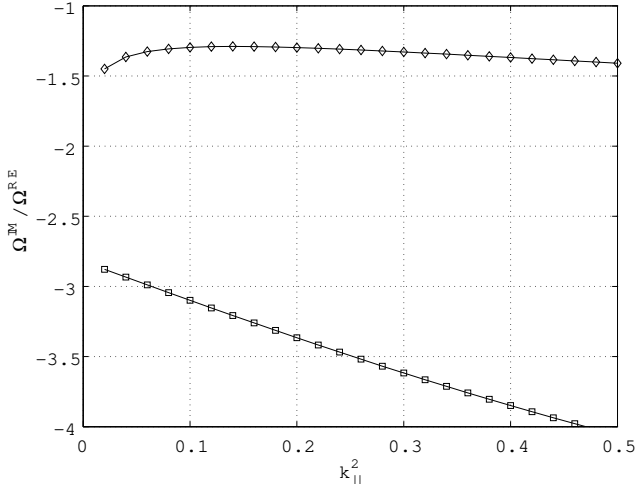}
  \caption{The zonal flow quality ($\Omega^{IM}/\Omega^{RE}$) as a function of $k_{\parallel}$ with $\epsilon_n = 0.0$ (diamond) and $\epsilon_n = 1.0$ (square) is shown. The other parameters are $\tau = 1$, $k_x = k_y = q_x = 0.3$, $\mu = 0$ and $\eta_i = 4.0$.}
\end{figure}

The zonal flow growth rate (positive) and real frequency (negative) (normalized to $c_s/L_n$) as a function of $k_{\parallel}$ with $\epsilon_n = 0.0$ (diamond) and $\epsilon_n = 1.0$ (square) are displayed in Figure 5. The other parameters are $\tau = 1$, $k_x = k_y = q_x = 0.3$, $\mu = 0$ and $\eta_i = 4.0$. A rapidly increasing zonal flow growth rate and real frequency is found for small parallel wave numbers and this result is corroborated with the approximate Eqs. 24 - 25 of the change in zonal flow $\Omega$. This also shows that the slab like generated zonal flows are driven by the acoustic waves through the ITG mode. This is the reason for the comparatively larger real frequency of the slab like driven zonal flow. The growth rate and real frequency for the toroidally generated zonal flows are only weakly dependent on the parallel wave number.

\begin{figure}
  \includegraphics[height=.3\textheight]{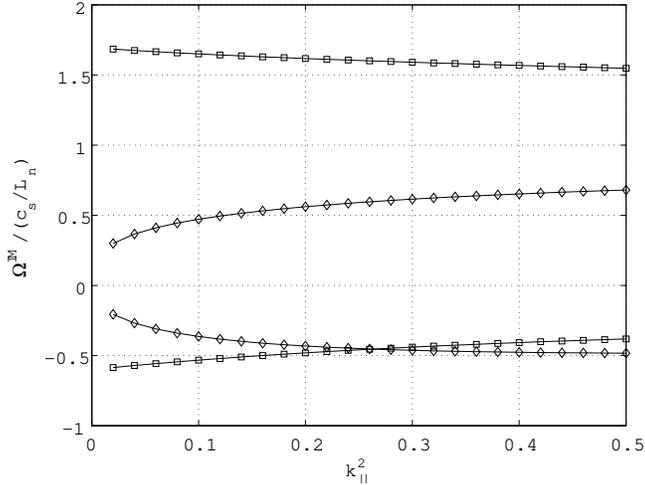}
  \caption{The zonal flow growth rate (positive) and real frequency (negative) (normalized to $c_s/L_n$) as a function of $k_{\parallel}$ with $\epsilon_n = 0.0$ (diamond) and $\epsilon_n = 1.0$ (square) are displayed. The other parameters are $\tau = 1$, $k_x = k_y = q_x = 0.3$, $\mu = 0$ and $\eta_i = 4.0$.}
\end{figure}

\section{Summary}
In this paper, a fifth order algebraic dispersion relation for zonal flow generation is derived including the effects of parallel ion motion. The emphasis is on investigating the difference in of zonal flows generated from slab like and toroidal Ion-Temperature-Gradient (ITG) mode turbulence. This is of special interest since it has been found recently that the suppression of turbulence by quasi stationary zonal flows are weak in linear experiments while it is expected that they significantly contribute to turbulence suppression in toroidal devices.

The physical model for the ITG mode are the equations for continuity, temperature and parallel ion motion. The electrons are considered to be Boltzmann distributed and the system of equations is closed by the quasi neutrality condition. The secondary generated zonal flow time evolution is governed by a Hasegawa-Mima like equation. The zonal flow excitation is studied through coupling of coherent modes. The drift wave is considered to be a monochromatic pump wave which may couple to the zonal flow through symmetric sidebands, giving a system of four coupled waves. The corresponding system may be reduced to a fifth order dispersion relation for the zonal flow. 

In this study the results are also interpreted in terms of quality of zonal flows i.e. the ratio of zonal flow growth rate and real frequency ($\Omega^{IM}/\Omega^{RE}$). The important finding is that the toroidally driven zonal flows exhibit a much larger ratio of growth rate and frequency compared to the slab like driven (factor of three difference), i.e. they have a higher quality. In most cases in scalings with $\eta_i$ the zonal flow growth rate and frequency follow the same qualitative behavior as the linear growth rate and real frequency. More importantly the quality of zonal flows is therefore rather insensitive to a scaling in $\eta_i$, in contrast to the results reported using the Wave Kinetic Equation (WKE) modeling. In many aspects the results found comparing the WKE modeling and the coherent mode coupling are qualitatively similar. In the WKE modeling a strongly decreasing zonal flow quality with increasing $\eta_i$ was found, however in the present model the decrease of zonal flow quality is visible for larger $\eta_i$ (not shown explicitly in this paper). In all parameter scalings the coherent mode coupling model agrees well with a model derived using the WKE model for the same background physics.

It is also shown that the slab like generated zonal flows are driven by the acoustic waves through the ITG mode. This is the reason for the comparatively larger real frequency of the slab like driven zonal flow. The growth rate and real frequency for the toroidally generated zonal flows are only weakly dependent on the parallel wave number.

To this end, in terms of quality, the toroidally driven zonal flows have higher values compared to the slab like generated zonal flows such as those found in CLM. The slab like generated zonal flows are inherently more oscillating zonal flows and hence the turbulence suppression found is much smaller.~\cite{a19} It should also be noted that the toroidally generated zonal flows exhibit large values of the growth rate ($\Omega^{IM}/\gamma_{ITG}>1$) compared to the slab case. All these results support the findings that the turbulence suppression is smaller in linear devices with small curvature (in the present paper represented with small $\epsilon_n$). There are still some unknown factors e.g. the effects of anisotropic ion temperature and temperature gradient that may shift the result. This will be the subject of a future paper.

\section{Acknowledgment}
This research was supported by Japan Society for the Promotion of Science (JSPS). The authors are also indebted to Mr. J. Douglas and Miss K. Burke for proofreading the manuscript.

\end{document}